# Approximating resonances with the Complex Absorbing Potential Method

Plamen Stefanov[*]
Department of Mathematics
Purdue University
West Lafayette, IN 47907, USA

**Abstract**

We study the Complex Absorbing Potential (CAP) Method in computing quantum resonances of width $c(h) = O(h^N)$, $N \gg 1$. We show that up to $h^{-M}\sqrt{c(h)} + O(h^\infty)$ error, $M \gg 1$, resonances are perturbed eigenvalues of the CAP Hamiltonian $P(h) - \mathrm{i}W$, and vice versa, where $W$ is the CAP with non-negative real part supported outside the trapping region. In some cases, the error terms are exponentially small.

## 1 Introduction

The purpose of this paper is to try to justify mathematically the *Complex Absorbing Potential* (CAP) method in computing quantum resonances. Let $P(h) = -h^2\Delta + V$ be the semiclassical Schrödinger operator with compactly supported potential $V(x)$ (in fact, we work with more general "black-box" Hamiltonians). Quantum resonances in a neighborhood $\Omega$ of some $E > 0$ are defined as the poles of the meromorphic extension of the resolvent $(P(h) - z)^{-1}$ from $\{\operatorname{Im} z > 0\} \cap \Omega$ to $\Omega$. They can also be defined as the eigenvalues of the complex-scaled version $P_\theta(h)$ of $P(h)$ (see section 7). We refer to [Z2] for a general introduction into resonance theory. In chemistry, resonances appear as metastable states. In this paper, we are interested in approximating resonances with $-\operatorname{Im} z = O(h^N)$, $N \gg 1$. Such resonances may exist only if $P(h)$ is trapping for the energy levels considered. A typical example is a potential well, another example are Hamiltonians with an elliptic periodic ray.

Since the interaction occurs only near $\operatorname{supp} V$, for numerical computations, the dynamics for large $|x|$ should not matter. One the other hand, working in unbounded domain is inconvenient. One way to "model infinity" is to use the complex scaled $P_\theta$, and impose Dirichlet conditions on a large sphere, placed behind the region where complex scaling occurs. The latter has been used in numerical computation of resonances, see [LZ]. Another way is the CAP method, which has the advantage of perturbing $P(h)$ by a zero order term, is to add to the latter a potential $-\mathrm{i}W(x)$ with $W \geq 0$, that is supported outside $\operatorname{supp} V$. The underlying idea is that $-\mathrm{i}W$ absorbs the signals without reflecting them (up to $O(h^\infty)$ error). Then one can impose Dirichlet or other boundary conditions on a large sphere encompassing $\partial(\operatorname{supp} W)$, and this should not create new reflections up to $O(h^\infty)$ terms. The CAP method was developed in [SM], [SCM] for various quantum computations and gives very good results. It has been used widely after that, see also [NM], [VH] and the references there. In the context of approximating resonances, the CAP method has been used in [RM], [PC], [MN].

---

[*]Partly supported by NSF Grant DMS-0400869

In this paper, we show that in a neighborhood of the real axis of polynomial width $c(h) = O(h^N)$, the eigenvalues of $Q(h) := P(h) - iW$ are perturbed resonances, and the resonances are perturbed eigenvalues of $Q(h)$. The error, up to a fixed polynomial factor, is $\max\{\sqrt{c(h)}, e^{-h^{-2/3}+\varepsilon}\}$. It would be interesting to know whether a suitable choice of $W$ would allow us to replace the latter exponential by $e^{-C/h}$. We also allow supp $V$ and supp $W$ to intersect as long as $P(h)$ is non-trapping on supp $W$. This introduces an $O(h^\infty)$ error however.

**Acknowledgments:** The author would like to thank Maciej Zworski for proposing this problem, and for useful discussions and remarks.

## 2 Main results

We define first an auxiliary operator $P_0(h)$ that represents $P(h)$ for large $|x|$. Fix $0 < R_0 < R_0'$. Let $P_0(h) = \sum_{|\alpha| \leq 2} a_\alpha(x)(hD)^\alpha$ be a formally self-adjoint operator that is a compactly supported perturbation of $-h^2 \Delta$ in $\mathbf{R}^n$, i.e.,

$$P_0(h) = -h^2 \Delta \quad \text{for } |x| \geq R_0'. \tag{1}$$

Assume that $P_0(h)$ is classically elliptic (i.e., $\sum_{|\alpha|=2} \xi^\alpha \neq 0$ for $\xi \neq 0$) with smooth coefficients. Here and below, we denote various positive constants by $C$. Fix $0 < a_0 < b_0 < \infty$. In what follows, we are always going to work with energy levels $E$ included in $[a_0, b_0]$. Assume also that $P_0(h)$ is non-trapping for such energy levels. The latter means the following: let $p_0(x, \xi) = \sum_{|\alpha| \leq 2} \xi^\alpha$ be the semiclassical symbol of $P_0(h)$. Then we require that for any $(x, \xi) \in T^*\mathbf{R}^n$ with $a_0 \leq p_0(x, \xi) \leq b_0$, we have that $|\Phi^t(x, \xi)| \to \infty$, as $t \to \infty$, where $\Phi^t$ is the Hamiltonian flow associated with $p_0$.

Let $P_0(h)$ be an operator satisfying the black box assumptions in a Hilbert space $\mathcal{H}$ described in section 3. The black box is included in the ball $B(0, R_0)$. We require that

$$P(h) = P_0(h) \quad \text{for } |x| > R_0. \tag{2}$$

We consider two CAP operators: one, that we denote by $Q_\infty(h)$ acts in the unbounded space, and the other one, denoted by $Q_R(h)$ acts in a domain obtained from the original one by restricting to the ball $B(0, R)$, $R \gg 0$, and imposing Dirichlet boundary conditions (Neumann b. c. would work equally well).

Let $W \in L^\infty$ be a complex-valued potential such that

$$\operatorname{Re} W(x) \geq 0, \quad \operatorname{supp} W \subset \mathbf{R}^n \setminus B(0, R_1), \quad R_0 < R_1, \tag{3}$$

We also assume also that for some $\delta_0 > 0$, $R_2 > R_1$,

$$\operatorname{Re} W \geq \delta_0 \quad \text{for } |x| > R_2. \tag{4}$$

And finally, we require that

$$|\operatorname{Im} W| \leq C(\operatorname{Re} W)^{1/2}. \tag{5}$$

This condition is quite reasonable: it means that Im $W$, which contributes a real term in $-iW$ and can reflect signals, has to be dominated by the absorbing part Re $W$ in the sense given above. This condition is certainly satisfied if $W$ is real. Set

$$Q_\infty(h) = P(h) - iW \quad \text{in } \mathcal{H}. \tag{6}$$

Given $R > R_2$, let $\mathcal{H}_R$ be as in section 3 (roughly speaking, it is the restriction of $\mathcal{H}$ on the ball $B(0, R)$), and let $P_R(h)$ be the Dirichlet realization of $P(h)$ there. Set

$$Q_R(h) = P_R(h) - iW \quad \text{in } \mathcal{H}_R. \tag{7}$$

Clearly, $Q_\infty(h)$ and $Q_R(h)$ are closed unbounded operators with $\mathcal{D}(Q_\infty(h)) = \mathcal{D}(P(h))$, $\mathcal{D}(Q_R(h)) = \mathcal{D}(P_R(h))$ and $\operatorname{Im} z > 0$ belongs to its resolvent sets. We prove in Proposition 1 that for any $h > 0$, the spectrum of $Q_\infty(h)$ in $\operatorname{Im} z > -\delta_0$ consists only of eigenvalues of finite multiplicities. The same is true for $Q_R(h)$ without the restriction $\operatorname{Im} z > -\delta_0$. Note that in most interesting situations, $P(h)$ has no positive real eigenvalues, then Propositions 1, 4 imply that the same holds for $Q_\infty(h)$, $Q_R(h)$.

Note that we did not assume that $R'_0 < R_1$. We allow $W$ to start rising in the region where $P(h)$ is still not equal to $-h^2\Delta$ and may not have analytic coefficients (so complex scaling is impossible there) but is non-trapping. Such an example is shown in Figure 1 below. On the other hand, if $R'_0 < R_1$, or more generally, if $P(h)$ has analytic coefficients in a neighborhood of supp $W$, then we can improve the estimates on the "resolution" of the CAP method from $O(h^\infty)$ to exponentially small, see the theorems below.

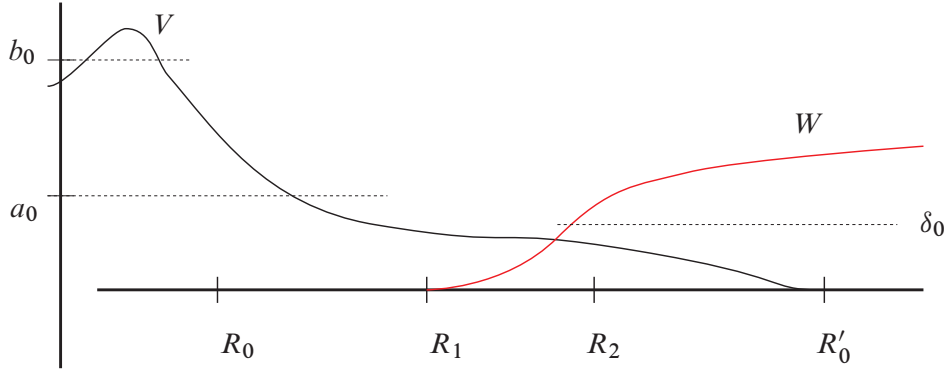

Figure 1: Sketch of a typical $V$ and $W$ in the case $P(h) = -h^2\Delta + V(x)$, $R_1 < R'_0$.

Our first result estimates the distance between Res $P(h)$ and Spec $Q(h)$, where $Q(h)$ is either $Q_\infty(h)$, or $Q_R(h)$, if we stay close to the real axis, but does not give information about the number of resonances/eigenvalues close to each other or about their multiplicities. The latter is addressed in Theorem 2. Theorem 1 can be considered as a partial case of Theorem 2 below with improved error however. The reason we formulate it separately, besides the improved error estimate, is that its proof is much more transparent, see section 6.

Note that in most interesting situations, including that of the Schrödinger operator, the number $n^\sharp$ introduced in section 3 is simply equal to the dimension $n$.

**Theorem 1** *Assume that $h \in H$, where $H \subset (0, 1]$, and $0$ is an accumulation point of $H$. Let $Q(h)$ denote either $Q_\infty(h)$, or $Q_R(h)$.*

*(a) Assume that $R'_0 < R_1$. Let $z_0(h)$ be a resonance in*

$$[a_0, b_0] + i\left[-\left(h^{n^\sharp+1}/C \log \frac{1}{h}\right)^2, 0\right], \quad C \gg 1, \tag{8}$$

where $0 < a_0 < b_0 < \infty$. Then for $H \ni h \ll 1$, there exists an eigenvalue of $Q(h)$ in

$$\left[\operatorname{Re} z_0(h) - \varepsilon(h) \log \frac{1}{h}, \operatorname{Re} z_0(h) + \varepsilon(h) \log \frac{1}{h}\right] + i[-\varepsilon(h), 0], \tag{9}$$

where $\varepsilon(h) = C_1 h^{-n^\sharp - 1/2} \sqrt{-\operatorname{Im} z_0(h)} + e^{-\gamma(R_1)/h}$. The constant $\gamma(R_1) > 0$ satisfies $\gamma(R_1) \geq (R_1 - R_0')/C_0$ and $C_1 > 0$ can be chosen uniform if $R_1$ belongs to a bounded interval.

(b) Assume that $R_1 \leq R_0'$. Then (a) holds with $\varepsilon(h) = C_1 h^{-n^\sharp - 1/2} \sqrt{-\operatorname{Im} z_0(h)} + O(h^\infty)$.

(c) Let $w_0(h)$ be an eigenvalue of $Q(h)$ in (8). Fix $B > 0$. Then for $H \ni h \ll 1$, $P(h)$ has a resonance in

$$\left[\operatorname{Re} w_0(h) - \delta(h) \log \frac{1}{h}, \operatorname{Re} w_0(h) + \delta(h) \log \frac{1}{h}\right] + i[-\delta(h), 0], \tag{10}$$

where $\delta(h) = C_2 B h^{-n^\sharp - 1} \sqrt{-\operatorname{Im} w_0(h)} + e^{-B/h}$.

**Remark.** For a large class of operators $P(h)$, including the Schrödinger operator $P(h) = -h^2 \Delta + V(x)$, $V \in C_0^\infty$ (and many more), Burq [B1], [B2] proved that for any $0 < a_0' < b_0' < \infty$, $\exists C > 0$, such that for $h \ll 1$,

$$\left([a_0', b_0'] + i\left[-e^{-C/h}, 0\right]\right) \cap \operatorname{Res} P(h) = \emptyset. \tag{11}$$

Then one can choose $R_1 \gg 1$ so that the exponential term in $\varepsilon(h)$ is absorbed by the first one. In other words, we have to push the absorbing region far enough to eliminate the exponential error term. Similarly, the $e^{-B/h}$ in $\delta(h)$ in (c) will be accumulated by the first one.

In particular, Theorem 1 implies, that if $Q(h)$ has an eigenvalue $w(h)$ with $-\operatorname{Im} w_0(h) = e^{-\alpha(h)/h}$, $1/C \leq \alpha(h) \leq C$, then there is a resonance $z(h)$ with $-\operatorname{Im} z = e^{-\beta(h)/h}$, where $\alpha(h)/2 - O(h \log(1/h)) \leq \beta(h) \leq 2\alpha(h) + O(h \log(1/h))$, provided that $R_1 \gg 1$. The latter condition is nod needed for the first inequality.

Given $\Omega \subset \mathbb{C}$, let $N_P(\Omega)$ and $N_Q(\Omega)$ denote the number of resonances of $P(h)$, and respectively the eigenvalues of $Q(h)$, in $\Omega$, counted with their multiplicities. Next theorem allows us to estimate the number of resonances in a box close to the real axis, by the number of eigenvalue of $Q(h)$.

**Theorem 2** *Let $H$ be as in Theorem 1. Let $Q(h)$ denote either $Q_\infty(h)$, or $Q_R(h)$. Fix $0 < a_0 < b_0 < \infty$, and let*

$$\Omega(h) = [a(h), b(h)] + i[-c(h), 0], \tag{12}$$

*where $a_0 \leq a(h) < b(h) \leq b_0$, $0 < e^{-h^{-2/3+\varepsilon_0}} \leq c(h) \leq h^M$, and $b(h) - a(h) \geq 2c(h)$.*

*(a) Assume that $R_0' < R_1$. Then there exist $N > 0$, $M > 0$, such that*

$$N_Q(\Omega_-(h)) \leq N_P(\Omega(h)) \leq N_Q(\Omega_+(h)), \tag{13}$$

*where*

$$\begin{aligned}
\Omega_- &= \left[a(h) + c(h), b(h) - c(h)\right] + i\left[-h^N c^2(h), 0\right], \\
\Omega_+ &= \left[a(h) - h^{-N} \sqrt{c(h)}, b(h) + h^{-N} \sqrt{c(h)}\right] + i\left[-h^{-N} \sqrt{c(h)}, 0\right].
\end{aligned}$$

*Moreover, the first inequality in (13) holds under the assumption of the weaker lower bound for $c(h)$: $e^{-C/h} \leq c(h)$.*

*(b) Assume that $R_1 \leq R_0'$. Then (a) hold with $\Omega_+$ replaced by*

$$\Omega_+ = \left[a(h) - h^{-N}\sqrt{c(h)} - O(h^\infty), b(h) + h^{-N}\sqrt{c(h)} + O(h^\infty)\right] + i\left[-h^{-N}\sqrt{c(h)} - O(h^\infty), 0\right].$$

**Remark.** We will note without a proof that the $O(h^\infty)$ terms in Theorem 1(b) and Theorem 2(b) can be replaced by $e^{-1/Ch}$ if the coefficients of $P(h)$ are analytic for $|x| \geq R_1 - \varepsilon$, $0 < \varepsilon \ll 1$. On the other hand, in this case, one can perform complex scaling for $|x| > R_1$.

## 3 Black Box assumptions

We work in the general framework of *black-box scattering* proposed by Sjöstrand and Zworski [SjZ1] (see also [Sj1], [TZ1]). We consider only compactly supported perturbations of the semiclassical Schrödinger operator $-h^2\Delta$. Let $\mathcal{H}$ be a complex Hilbert space of the form

$$\mathcal{H} = \mathcal{H}_{R_0} \oplus L^2(\mathbf{R}^n \setminus B(0, R_0)),$$

where $R_0 > 0$ is fixed and $B(0, R_0)$ is the ball centered at the origin with radius $R_0$. We consider a family of self-adjoint unbounded operators $P(h)$ in $\mathcal{H}$ with common domain $\mathcal{D}$, whose projection onto $L^2(\mathbf{R}^n \setminus B(0, R_0))$ is $H^2(\mathbf{R}^n \setminus B(0, R_0))$. In what follows we will denote by $\mathbf{1}_{B(0,R_0)}$ the orthogonal projector onto $\mathcal{H}_{R_0}$. We will also denote the same projector by $\mathbf{1}_{|x| \leq R_0}$, and will use the notation $\mathcal{H}_R$ for the space $\mathcal{H}_{R_0} \oplus L^2(B(R, 0) \setminus B(R_0, 0))$, where $R > R_0$. We assume that

$$\mathbf{1}_{B(0,R_0)}(P(h) + i)^{-1} : \mathcal{H} \to \mathcal{H}$$

is compact. Outside $\mathcal{H}_{R_0}$, $P(h)$ is assumed to coincide with $P_0(h)$, see (2), i.e.,

$$\mathbf{1}_{\mathbf{R}^n \setminus B(0,R_0)} P(h)u = P_0(h)\left(u|_{\mathbf{R}^n \setminus B(0,R_0)}\right).$$

For $|x| > R_0'$, we have $P(h) = -h^2\Delta$. Finally, we assume that $P(h) > -C_0$, $C_0 > 0$. Under those assumptions, one can define (the semi-classical) resonances $\operatorname{Res} P(h)$ of $P(h)$ in a conic neighborhood of the real axis by the method of complex scaling (see [SjZ1], [Sj1]). Resonances are also poles of the meromorphic continuation of the resolvent $(P(h) - z)^{-1} : \mathcal{H}_{\text{comp}} \to \mathcal{H}_{\text{loc}}$ from $\operatorname{Im} z > 0$ into a conic neighborhood of the real line. We will denote the so continued resolvent by $R(z, h)$.

As in [SjZ1], [Sj1], we construct a reference selfadjoint operator $P^\sharp(h)$ from $P(h)$ on $\mathcal{H}^\sharp = \mathcal{H}_{R_0'} \oplus L^2(M \setminus B(0, R_0'))$, where $M = (\mathbf{R}/R\mathbf{Z})^n$ for some $R \gg R_0'$. Then for the number of eigenvalues of $P^\sharp(h)$ in a given interval $[-\lambda, \lambda]$, we assume

$$\#\{z \in \operatorname{Spec} P^\sharp(h); \ -\lambda \leq z \leq \lambda; \} \leq C(\lambda/h^2)^{n^\sharp/2}, \quad \lambda \geq 1, \tag{14}$$

with some $n^\sharp \geq n$. In most interesting cases, including that of $P(h) = -h^2\Delta + V(x)$, we have $n = n^\sharp$. Estimate (14) implies (see [SjZ1] and [Sj1]) that

$$\#\{z \in \operatorname{Res} P(h); \ 0 < a_0 \leq \operatorname{Re} z \leq b_0; \ 0 \leq -\operatorname{Im} z \leq c_0\} \leq C(a_0, b_0, c_0) h^{-n^\sharp}. \tag{15}$$

Polynomial estimates of this type have been proved also in [M], [Z1], [SjZ1], [V1], [Sj1].

Finally, we recall an a priori estimate on the resolvent, see [TZ1] and the references there. For any precompact region $\Omega \subset \mathbf{C} \setminus \{0\}$, $\exists A > 0$, such that

$$\|\chi R(z, h)\chi\| \leq e^{Ah^{-n^\sharp} \log(1/g(h))} \quad \text{for } z \in \Omega, \ \operatorname{dist}(z, \operatorname{Res} P(h)) \geq g(h), \tag{16}$$

for any $0 < g(h) = o(h^{-n^\sharp})$.

In what follows, we denote by $C$ various positive constants that may change form line to line. With some abuse of notation, $\operatorname{supp} \chi \subset B(0, R)$ (the latter is the ball centered at 0 with radius $R$) actually means that $\chi = \mathbf{1}_{B(0,R_0)} + \chi'$, where $\operatorname{supp} \chi' \subset B(0, R)$, etc. We also use the notation $\chi_1 \prec \chi_2$ to indicate that $\chi_2 = 1$ in a neighborhood of $\operatorname{supp} \chi_1$. We will often suppress the dependence on $h$, i.e., we will denote $P(h)$ by $P$, etc., to simplify the notation.

## 4 Properties of $Q_\infty(h)$ and $Q_R(h)$

### 4.1 Analysis of $Q_\infty(h)$

**Proposition 1** *For any $h > 0$, the resolvent $(Q_\infty(h) - z)^{-1}$ extends meromorphically from $\{\operatorname{Im} z > 0\}$ into $\{\operatorname{Im} z > -\delta_0\}$. The poles of $(Q_\infty(h) - z)^{-1}$ are eigenvalues of $Q_\infty$ of finite multiplicity. Moreover, $0 \neq z \in \mathbf{R}$ is an eigenvalue of $Q_\infty(h)$ if and only if it is an eigenvalue of $P(h)$.*

*Proof:* The proof basically follows from the fact that for $\operatorname{Im} z > -\delta_0$, $Q_\infty - z$ is a relatively compact perturbation of the invertible operator $P - \mathrm{i} W_1 - z$, where $W_1 := \delta_0$ for $|x| < R_2$, $W_1 := W$ otherwise.

More precisely, define the following candidate for an approximate right inverse of $Q_\infty - z$. Let $\chi_1 + \chi_2 + \chi_3 = 1$ be a smooth partition of unity, such that $\chi_1 = 1$ near $B(0, R_0)$, $\operatorname{supp} \chi_1 \subset B(0, (R_0 + R_1)/2)$; $\operatorname{supp} \chi_3 \subset \mathbf{R}^n \setminus B(0, R_2)$, and $\chi_3 = 1$ for $|x| \gg 1$. Let $\tilde{\chi}_i \succ \chi_i$, $i = 1, 2, 3$ and have the same support properties but the sum does not equal 1. Let $\operatorname{Im} z_0 > 0$ and set

$$E(z) = \tilde{\chi}_1 (P - z_0)^{-1} \chi_1 + \tilde{\chi}_2 (P_0 - \mathrm{i} W - z_0)^{-1} \chi_2 + \tilde{\chi}_3 (P_0 - \mathrm{i} W_1 - z)^{-1} \chi_3,$$

where $W_1$ is as above, and in particular, $W_1 = W$ for $|x| > R_2$, and $\operatorname{Re} W_1 \geq \delta_0$, see (4). The latter inequality implies that $(P_0 - \mathrm{i} W_1 - z)^{-1}$ is holomorphic for $\operatorname{Im} z > -\delta_0$.

Apply $Q_\infty - z$ to $E(z)$ to get

$$\begin{aligned}
(Q_\infty - z) E(z) &= [P_0, \tilde{\chi}_1](P - z_0)^{-1} \chi_1 + \tilde{\chi}_1 \big[ I + (z_0 - z)(P - z_0)^{-1} \big] \chi_1 \\
&\quad + [P_0, \tilde{\chi}_2](P_0 - \mathrm{i} W - z_0)^{-1} \chi_2 + \tilde{\chi}_2 \big[ I + (z_0 - z)(P_0 - \mathrm{i} W - z_0)^{-1} \big] \chi_2 \\
&\quad + [P_0, \tilde{\chi}_3](P_0 - \mathrm{i} W_1 - z)^{-1} \chi_3 + \chi_3.
\end{aligned} \tag{17}$$

Therefore,

$$(Q_\infty - z) E(z) = I + K(z), \tag{18}$$

where

$$\begin{aligned}
K(z) &= [P_0, \tilde{\chi}_1](P - z_0)^{-1} \chi_1 + (z_0 - z)\tilde{\chi}_1 (P - z_0)^{-1} \chi_1 \\
&\quad + [P_0, \tilde{\chi}_2](P_0 - \mathrm{i} W - z_0)^{-1} \chi_2 + (z - z_0)\tilde{\chi}_2 (P_0 - \mathrm{i} W - z_0)^{-1} \chi_2 \\
&\quad + [P_0, \tilde{\chi}_3](P_0 - \mathrm{i} W_1 - z)^{-1} \chi_3 \\
&= K_1 + K_2(z) + K_3 + K_4(z) + K_5(z).
\end{aligned} \tag{19}$$

Clearly, $K(z)$ is a compact operator, depending analytically on $z \in \{\operatorname{Im} z > -\delta_0\}$. We claim that for $\operatorname{Im} z_0 \gg 1$, and $z$ close to $z_0$, $\|K(z)\| \leq 1/2$, therefore $I + K(z)$ is invertible there. This follows form the following: by the spectral theorem, $\|(P - z_0)^{-1}\| \leq 1/\operatorname{Im} z_0$ for $\operatorname{Im} z_0 > 0$. This easily implies that $\|P(P - z_0)^{-1}\| \leq C$ uniformly in $z_0$ and $h$, if $\operatorname{Im} z_0 \geq 1$, and $\operatorname{Re} z_0$ is bounded. By standard semi-classical elliptic estimates, we get that for any $\chi \in C_0^\infty(\mathbf{R}^n \setminus B(0, R_0))$, $\|h^2 \Delta \chi (P - z_0)^{-1}\| \leq C$. Using the Fourier transform, we obtain $\|h\nabla\chi(P-z_0)^{-1}\| \le C/\sqrt{\operatorname{Im} z_0}$ under the same assumptions on $z_0$. This shows that for $\operatorname{Re} z_0$ fixed and $\operatorname{Im} z_0 \gg 1$, independent of $h \in (0,1]$, $\|K_1\| \le 1/10$. The proof for $K_3$ and $K_5$ is the same. This proves our claim.

Fix $z_0$, independent of $h \in (0,1]$, as above. By the analytic Fredholm theorem, $(I + K(z))^{-1}$ is meromorphic in $\{\operatorname{Im} z > -\delta_0\}$. Then $E(z)(I+K(z))^{-1}$ is a right inverse for $Q_\infty - z$. A left inverse is constructed in the same way by switching $\tilde\chi_i$ and $\chi_i$, $i = 1, 2, 3$ in (17), and this gives us in fact ${}^t E(z)$, where ${}^t E(z)$ is the transpose of $E$, i.e., the operator with Schwartz kernel obtained by switching the variables $x$ and $y$. Then the left and right inverses have the same poles, they coincide outside the poles, therefore, they are equal as meromorphic functions. Therefore,

$$(Q_\infty - z)^{-1} = E(z)(I + K(z))^{-1}. \tag{20}$$

Thus, the l.h.s. above is meromorphic in $\operatorname{Im} z > -\delta_0$, with poles among those of $(I+K(z))^{-1}$. Moreover, the residue of the resolvent $(Q_\infty - z)^{-1}$ at such a pole is of finite order and rank, because the same is true for the residue of $(I+K(z))^{-1}$. The second statement of the proposition follows from the general theory of non-selfadjoint operators.

Now, let $z$ be an eigenvalue of $Q_\infty$, and let $f$ be an eigenfunction corresponding to it. Then $0 = \operatorname{Im}((Q_\infty - z)f, f)$, which implies

$$-\operatorname{Im} z \|f\|^2 = ((\operatorname{Re} W)f, f). \tag{21}$$

If $z \in \mathbf{R}$, then $(\operatorname{Re} W)^{1/2} f = 0$. Then $Wf = 0$ as well, see (5), thus $Pf = zf$. On the other hand, if $z \ne 0$ is a real eigenvalue of $P$, then all corresponding eigenfunctions are supported in the "black box" [Sj3], therefore $z$ is an eigenvalue for $Q_\infty$ as well. $\square$

**Proposition 2** *Let $a < b$, $0 < c < \delta_0$, and $\Omega = [a,b] + \mathrm{i}[-c, 0]$. Then for the number $N_{Q_\infty}(\Omega)$ of eigenvalues of $Q_\infty$ in $\Omega$ we have*

$$N_{Q_\infty}(\Omega) \le C h^{-n^\sharp}, \tag{22}$$

*where $C$ depends on $\Omega$ and $Q_\infty$ only.*

*Proof.* We use the representation (18), (19), where $z_0$ with $\operatorname{Im} z_0 \gg 0$ is chosen as above, and in addition we can assume that for some $r > 0$, the disk $D(z_0, r)$ contains $\bar\Omega$ but its closure is included in $\operatorname{Im} z > -\delta_0$.

Recall that $\|K_1\| + \|K_3\| < 2/10$ for all $0 < h \le 1$. Arguing as in the previous section, we see that $\|K_5(z)\| = O(h)$. Note that $K_j = O(h^\infty)$, $j = 1, 3, 5$, if $W$ is smooth. Therefore, one can write

$$1 + K(z) = \bigl(1 + \tilde K(z)\bigr)\bigl(K_1 + K_3 + K_5(z)\bigr), \quad \tilde K(z) := \bigl(K_2(z) + K_4(z)\bigr)\bigl(K_1 + K_3 + K_5(z)\bigr)^{-1}.$$

We will estimate the function

$$f(z) = \det\bigl(I + \tilde K^{n^\sharp}(z)\bigr), \quad z \in \Omega. \tag{23}$$

To this end, it suffices to estimate the characteristic values $\mu_j(K_2(z))$, $\mu_j(K_4(z))$ for $z \in \Omega$.

We estimate $\mu_j(K_2)$ first. It is known [Sj3, sec. 6] that (14) implies the estimate

$$\mu_j(\tilde\chi_1(P-z_0)^{-1}\chi_1) \le C\bigl(1 + h^2 j^{2/n^\sharp}\bigr)^{-1}, \tag{24}$$

since the same holds for the characteristic values of $(P^\sharp - z_0)^{-1}$. This implies the same kind of estimate for $\mu_j(K_2)$

$$\mu_j(K_2) \le C\bigl(1 + h^2 j^{2/n^\sharp}\bigr)^{-1}. \tag{25}$$

To estimate $\mu_j(K_4)$, denote by $P_0^\sharp$ the operator defined similarly to $P^\sharp$ in section 3, but obtained from $P_0$ instead from $P$. Note that

$$(P_0^\sharp - i)^{-1}(P_0^\sharp - i)\tilde{\chi}_2(-h^2\Delta - iW - z_0)^{-1}\chi_2 = (P_0^\sharp - i)^{-1}L,$$

where $\|L\| = O(1)$. Using the inequalities $\mu_j(AB) \leq \|A\|\mu_j(B)$, $\mu_j(AB) \leq \|B\|\mu_j(A)$, the problem is then reduced to estimates of the characteristic values of $(P_0^\sharp - i)^{-1}$, and they satisfy (25) by the well known Weyl type semiclassical asymptotics, see [DSj]. Therefore,

$$\mu_j(K_4) \leq C\left(1 + h^2 j^{2/n^\sharp}\right)^{-1}. \tag{26}$$

Therefore, using Fan's inequality $\mu_{i+j-1}(A+B) \leq \mu_i(A) + \mu_j(B)$, we get that $\mu_j(K)$ satisfies (26) as well.

To estimate $\mu_j(\tilde{K}^{n^\sharp}(z))$, we use another well-known inequality $\mu_{i+j-1}(AB) \leq \mu_i(A)\mu_j(B)$, iterated $n^\sharp$ times, to get that

$$\mu_j\left(\tilde{K}^{n^\sharp}(z)\right) \leq C\left(1 + h^2 j^{2/n^\sharp}\right)^{-n^\sharp} \leq C\left(1 + jh^{n^\sharp}\right)^{-2}. \tag{27}$$

This yields $\sum \mu_j\left(\tilde{K}^{n^\sharp}(z)\right) = O(h^{-n^\sharp})$, and by well known estimates, $|f(z)|$ is bounded by this sum, i.e.,

$$|f(z)| \leq Ch^{-n^\sharp}. \tag{28}$$

On the other hand, we have $f(z_0) = 1$. Thus by Jensen's inequality in $B(z_0, r + \varepsilon)$, $\varepsilon \ll 1$, the number of zeros of $f(z)$ in $D(z_0, r)$, and therefore in $\Omega$, is $O(h^{-n^\sharp})$. Those zeroes include the eigenvalues of $Q_\infty$, together with multiplicities, see e.g., [Sj3, Proposition 5.16]. This proves the proposition. $\square$

Next, we show that (16) holds for the resolvent of $Q_\infty(h)$ as well.

**Proposition 3** *Let $\Omega$ be as in Proposition 2. Then there exists $A = A(\Omega)$, such that*

$$\|(Q_\infty(h) - z)^{-1}\| \leq e^{Ah^{-n^\sharp}\log(1/g(h))} \quad \text{for } z \in \Omega, \operatorname{dist}(z, \operatorname{Spec} Q_\infty(h)) \geq g(h), \tag{29}$$

*for any $0 < g(h) = o(h^{-n^\sharp})$.*

*Proof*: The proposition follows from (28) and $f(z_0) = 1$ as in [PZ, sec. 4]. $\square$

Finally, let us mention that the inequality $-\operatorname{Im}((Q_\infty - z)f, f) = ((\operatorname{Re} W)f, f) + \operatorname{Im} z\|f\|^2 \geq \operatorname{Im} z\|f\|^2$, where $f \in \mathcal{D}(Q_\infty)$, implies

$$\|(Q_\infty - z)^{-1}\| \leq \frac{1}{\operatorname{Im} z}, \quad \operatorname{Im} z > 0. \tag{30}$$

## 4.2 Analysis of $Q_R(h)$

We show next that all the properties of $Q(h)$ are preserved if we replace it by $Q_R(h)$, where $Q_R(h)$ is defined in (7). For the resolvent we then have

$$(Q_R(h) - z)^{-1} = (P_R(h) - z_0)^{-1}(I + K(z))^{-1},$$

where
$$K(z) = (-\mathrm{i}W - z + z_0)(P_R(h) - z_0)^{-1}.$$

and $\operatorname{Im} z_0 > 0$. For $\operatorname{Im} z_0 \gg 0$ and $z$ close to $z_0$, $\mathrm{I} + K(z)$ is invertible, therefore $(\mathrm{I} + K(z))^{-1}$ is a meromorphic family. The eigenvalues of $Q_R(h)$ can then be characterized as the poles of $(\mathrm{I} + K(z))^{-1}$. For each eigenvalue $z$, and eigenfunction $f$, (21) is still true, therefore if $z$ is real, then $Wf = 0$, so $f$ vanishes near $\partial B(0, R)$, and is also an eigenfunction of $P(h)$ as well. Similarly, if a non-zero real $z$ is an eigenvalue of $P(h)$, then it is an eigenvalue of $Q_R(h)$ as well. Thus Proposition 1 still holds for $Q_R(h)$.

Propositions 2 and 3 hold for $Q_R(h)$ as well. Indeed, as above, we need to prove (28), where $f(z)$ is as in (23) but with $\tilde{K}(z)$ replaced by $K(z)$ above. This can be done by estimating the characteristic values $\mu_j(K(z))$. They satisfy (25), (26) because the eigenvalues of $(P_R(h) - z_0)^{-1}$ satisfy them as well, and the latter follows from (14) as in (24). Let us summarize this in the following.

**Proposition 4** *For any $h > 0$, the resolvent $(Q_R(h) - z)^{-1}$ is a meromorphic function of $z \in \mathbf{C}$, and its poles are eigenvalues of $Q_R(h)$ of finite multiplicities. The last statement of Proposition 1, as well as Propositions 2, 3, and estimate (30) hold with $Q_\infty(h)$ replaced by $Q_R(h)$ as well.*

## 5  From quasimodes to resonances, revisited

In this section, we will review the connection between quasimodes and resonances developed in [StV], [TZ1], [St1], by improving some details. The first improvement is to formulate the theorem below for long range perturbations of the Laplacian, which does not require new efforts, see [St1] and [Sj3]. The second one is to formulate the result so that it would give resonances exponentially close to the quasimodes (not only the imaginary, but the real part as well), if the error is exponentially small. This is not new either, and follows from the recent versions of the lemma below, see e.g., [St2] but the corresponding implications to the resonances and quasimodes connection have not been formulated clearly so far, except for Remark 5 in [St1] that can be improved as well. And finally, asymptotic orthogonality of the quasimodes can be relaxed, it can be replaced by a linear independence stable under perturbations (40), see also [St4].

Next lemma is sometimes referred to as the "semiclassical maximum principle" [TZ1], see also [StV]. The version presented here is close to that in [St2].

**Lemma 1** *Let $0 < h < 1$ and $a(h) \leq b(h)$. Suppose that $F(z, h)$ is a holomorphic function of $z$ defined in a neighborhood of*
$$\Omega(h) = [a(h) - w(h), b(h) + w(h)] + \mathrm{i}[-\alpha(h)S_-(h), S_+(h)],$$

*where $0 < S_+(h) \leq S_-(h)$, $1 \leq \alpha(h)$, and $S_-(h)\alpha(h) \log \alpha(h) \leq w(h)$. If $F(z, h)$ satisfies*

$$|F(z, h)| \leq e^{\alpha(h)} \quad \text{on } \Omega(h), \tag{31}$$
$$|F(z, h)| \leq M(h) \quad \text{on } [a(h) - w(h), b(h) + w(h)] + \mathrm{i}S_+(h) \tag{32}$$

*with $M(h) \geq 1$, then there exists $h_1 = h_1(S_-, S_+, \alpha) > 0$ such that*

$$|F(z, h)| \leq e^3 M(h), \quad \forall z \in \tilde{\Omega} := [a(h), b(h)] + \mathrm{i}[-S_-(h), S_+(h)]$$

*for $h \leq h_1$.*

*Proof*: We follow [Sj3]. To simplify the notation, we will suppress the dependence on $h$. Set

$$f(z) = \log |F(z)| - \log M \frac{\operatorname{Im} z + \alpha S_-}{\alpha S_- + S_+} - \alpha \frac{S_+ - \operatorname{Im} z}{\alpha S_- + S_+}. \tag{33}$$

Then $f(z) = \operatorname{Re} \log F(z)$ is a subharmonic function near $\Omega$, and harmonic, if $F(z)$ has no zeros there. By (31), (32), $f \leq 0$ on the horizontal sides of $\Omega$. On the vertical sides, as well as anywhere in $\Omega$, $f$ satisfies $f(z) \leq \alpha$ by (31). Then

$$f(z) \leq \int_{\partial \Omega} \mathcal{P}(z, y) f(y)|_{\partial \Omega} \, dS_y, \tag{34}$$

where $\mathcal{P}(z, y)$ is the Poisson kernel in $\Omega$. Since $\Omega$ is a rectangle, one can use separation of variables to get an explicit expression of $\mathcal{P}(z, y)$, and the latter decays exponentially in "long domains" away from the vertical sides (see also [Sj2]). Therefore,

$$f(z) \leq \alpha e^{-C\omega/(\alpha S_- + S_+)} \quad \text{for } z \in \Omega, a \leq \operatorname{Re} z \leq b,$$

if $h \ll 1$, and it can be seen that $C$ can be any constant less than $\pi$, for example $C = 2$. Under our assumption for $\omega$, the inequality above implies that $f \leq 1$ for $z \in \Omega$, $a \leq \operatorname{Re} z \leq b$. If, in addition, $-\operatorname{Im} z \leq S_-$, then the last term in (33) is bounded from below by $-2$. Therefore,

$$\log |F(z)| \leq 1 + \log M + 2 \quad \text{in } \tilde{\Omega}(h),$$

which completes the proof. $\square$

The lemma still holds, if $F(h)$ is an operator valued function. Indeed, then one can apply the lemma to the function $(F(z, h)\phi, \psi)$ with $\|\phi\| = \|\psi\| = 1$ (then $h_1$ does not depend on $\phi, \psi$).

**Corollary 1** *Let $a(h) \leq b(h)$, and $F(z, h)$ be a holomorphic function near*

$$\Omega(h) = [a(h) - w(h), b(h) + w(h)] + i[-Ah^{-n^\sharp} \log \frac{1}{S(h)} S(h), S(h)], \tag{35}$$

*where $e^{-B/h} < S(h) < 1$, $B > 0$, and $2An^\sharp h^{-n^\sharp} \log \frac{1}{h} \log \frac{1}{S(h)} S(h) \leq w(h)$. If $F(z, h)$ satisfies*

$$|F(z, h)| \leq e^{Ah^{-n^\sharp} \log(1/S(h))} \quad \text{on } \Omega(h), \tag{36}$$

$$|F(z, h)| \leq \frac{1}{\operatorname{Im} z} \quad \text{on } \Omega(h) \cap \{\operatorname{Im} z > 0\}, \tag{37}$$

*then there exists $h_1 = h_1(S_-, S_+, \alpha) > 0$ such that for $h \leq h_1$,*

$$|F(z, h)| \leq \frac{e^3}{S(h)}, \quad \forall z \in \tilde{\Omega} := [a(h), b(h)] + i[-S(h), S(h)]. \tag{38}$$

*Proof*: We apply Lemma 1 with $\alpha(h) = Ah^{-n^\sharp} \log(1/S(h))$, $S_- = S_+ = S$, $M(h) = 1/S(h)$. $\square$

A typical application of the lemma is when one can find a quasimode, i.e., a quasiresonance $E(h) \in \mathbf{R}$ and a compactly supported $u(h)$ with the property $\|(P(h) - E(h))u(h)\| = R(h) = O(h^{-N})$, $N \gg 1$. Then we choose $S_-(h) = S_+(h) \sim R(h)$, $F(h) = \chi R(z, h)\chi$ with a suitable cut-off $\chi$, and then $M(h) =$ $1/S_+(h)$. The lemma then implies an existence of a resonance at a distance $O\big(R(h)h^{-n^\sharp-1}\log(1/h)\big)$ from $E(h)$. More details are given below. In some cases, one really needs $S_- \neq S_+$, see [St4].

We formulate next theorem for long-range perturbations of the Laplacian. We refer to [Sj1], [TZ1], [Sj3] for more details. We skip the definition of that latter, since we will apply the theorem below to short range perturbations only described in section 3.

**Theorem 3** *Let $h \in H \subset (0, h_0]$, and let $0$ be an accumulation point of $H$. Let $P(h)$ satisfy the long-range black box hypotheses. Let $0 < a_0 \leq a(h) \leq b(h) \leq b_0 < \infty$. Assume that for any $h \in H$, there exist $m(h) \in \{1, 2, \ldots\}$, $E_j(h) \in [a(h), b(h)]$, and $u_j(h) \in \mathcal{D}$, $\|u_j(h)\| = 1$, $1 \leq j \leq m(h)$, such that $\operatorname{supp} u_j(h) \subset K$, with $K$ a compact set in $\mathbf{R}^n$ independent of $h$, and the following properties are satisfied:*

$$\|(P(h) - E_j(h))u_j(h)\| \leq R(h), \tag{39}$$

$\forall \tilde{u}_j(h) \in \mathcal{H}$ with $\|\tilde{u}_j(h) - u_j(h)\| \leq h^N/M$, $1 \leq j \leq m(h)$, $\{\tilde{u}_j(h)\}_{j=1}^{m(h)}$ are linearly independent, 
$$\tag{40}$$

*where $R(h) \leq h^{n^\sharp + N + 1}/C \log(1/h)$, $C \gg 1$, $N \geq 0$, $M > 0$. Then there exists $C_0 = C_0(a_0, b_0, P) > 0$, such that for any $B > 0$, $\exists h_1 = h_1(A, B, M, N) \leq h_0$ such that for $H \ni h \leq h_1$, $P(h)$ has at least $m(h)$ resonances in*

$$\Big[a(h) - c(h)\log\frac{1}{h}, b(h) + c(h)\log\frac{1}{h}\Big] - \mathrm{i}[0, c(h)], \tag{41}$$

*where*

$$c(h) = \max\Big(C_0 B M R(h) h^{-n^\sharp - N - 1}, e^{-B/h}\Big).$$

**Remark.** As shown in [St1], if $u_j(h)$ are orthogonal, then (40) is fulfilled if $h^N/M < 1/m(h)$. Actually, the theorem implies that $m(h) = O(h^{-n^\sharp})$, so one can take $N = n^\sharp$, $M \gg 1$ in case of orthogonal quasimodes. If $|(u_i(h), u_j(h)) - \delta_{ij}| \leq \alpha/m(h)$, $\alpha < 1$, then this is still true, i.e., the conditions

$$|(u_i(h), u_j(h)) - \delta_{ij}| \leq \alpha/m(h), \quad h^N/M < \alpha/m(h), \quad 0 < \alpha < 1,$$

imply (40), and $h^N/M < \alpha/m(h)$ is always fulfilled for $N = n^\sharp$, $M \gg 1$.

*Proof of Theorem 3*: We will sketch the proof by pointing out the slight modifications needed in the proof of [St1, Theorem 1], see also [Sj3, Theorem 11.2].

Let $z_1(h), \ldots, z_{J(h)}(h)$ be all distinct resonances in

$$\Omega_2(h) := \big[a(h) - 2w(h), b(h) + 2w(h)\big] + \mathrm{i}\Big[-2Ah^{-n^\sharp}\log\frac{1}{S(h)}S(h), S(h)\Big].$$

(compare with (35)), where $0 < S(h) \ll 1$ and $w(h)$ will be specified below. Fix $\chi \in C_0^\infty$, $\chi \succ \mathbf{1}_{B(0, R_0')}$. The multiplicity of each $z_j(h)$ is given by the rank of the residue $A^{(j)}(h)$ of $\chi R(z, h)\chi$ at $z_j(h)$, see e.g. [SjZ1], [St1], and [Sj3] for the long-range case. We need to prove that $\tilde{m}(h) := \sum \operatorname{Rank} A^{(j)}(h) \geq m(h)$. Let $\Pi(h)$ be the orthogonal projection onto $\cup A^{(j)}(h)\mathcal{H}$, and let $\Pi'(h) = I - \Pi(h)$. Then $\operatorname{Rank} \Pi(h) \leq \tilde{m}(h)$, so it is enough to show that $\operatorname{Rank} \Pi(h) \geq m(h)$.

Analyzing the terms in the Laurent expansion of $\chi R(z, h)\chi$ at each resonance $z_j(h)$, it is proven in [St1], see also [Sj3], that $F(z, h) := \Pi'(h)\chi R(z, h)\chi$ is holomorphic in $\Omega(h)$, and satisfies $\|F(z, h)\| \leq 1/\operatorname{Im} z$ for $\operatorname{Im} z > 0$. It also satisfies (16), therefore (36) is fulfilled as long as $z \in [a_0/2, 2b_0] + \mathrm{i}[-1/C, 1/C]$, $C \gg 1$, and $\operatorname{dist}(z, \operatorname{Res} P(h)) \geq S(h)$.

Set
$$w(h) = 4n^\sharp A h^{-n^\sharp} \log \frac{1}{h} \log \frac{1}{S(h)} S(h),$$

and assume that $S(h)$ is such that $w(h) \leq a_0/2$. Since the diameter of the largest connected union of disks centered at resonances with radius $S(h)$, is $O(h^{-n^\sharp} S(h))$, as in [St1] we get that (36) is satisfied in $\Omega(h)$ given by (35). We can apply Corollary 1 to get (38). The existence of quasimodes however implies by (39),

$$\|\Pi'(h)u_j(h)\| \leq \|F(z,h)\| R(h) \leq \frac{e^3 R(h)}{S(h)}.$$

Therefore, for $\tilde{u}_j(h) = \Pi(h)u_j(h)$, we have $\|\tilde{u}_j(h) - u_j(h)\| \leq e^3 R(h)/S(h)$. If the latter does not exceed $h^N/M$, then $\tilde{u}_j(h)$ are linearly independent by (40), and the inequality Rank $\Pi(h) \geq m(h)$ follows. Therfore, we can choose $S(h) = \max(e^3 M h^{-N} R(h), e^{-2B/h})$ to get $\|\tilde{u}_j(h) - u_j(h)\| \leq h^N/M$.

Thus we have proved that there are at least $m(h)$ resonances in $\Omega(h)$. It can be easily seen that the domain (41) includes $\Omega(h)$ for $h \ll 1$, and that the assumptions on $R(h)$ imply the requirement on $w(h)$ above. □

The error estimate in the theorem can be improved if we take into account the contribution of one quasimode only (at the expense of losing information about multiplicities and clusters of resonances): if $E(h)$ is a real quasiresonance as in the theorem, then there exists a resonance $z(h)$, such that

$$|\operatorname{Re} z(h) - E(h)| \leq CR(h) h^{-n^\sharp - 1} \log \frac{1}{h}, \quad 0 \leq -\operatorname{Im} z(h) \leq CR(h) h^{-n^\sharp - 1}.$$

**Theorem 4** *The conclusions of Theorem 3 remain true with $P(h)$ replaced by $Q(h)$ and "resonances" replaced by "eigenvalues".*

*Proof*: The proof is the same as above. Instead of (15), (16), and the estimate $\|\chi R(z,h)\chi\| \leq 1/\operatorname{Im} z$, $\operatorname{Im} z > 0$, we use Propositions 2 and 3, and (30). □

## 6 Proof of Theorem 1

*Proof of (a):* Let $z_0(h)$ be a resonance in (8), and let $u(h)$ be a corresponding resonant state. Then by [B1], and Proposition 3 in [St3],

$$\int_{|x|=\rho} \left(|u|^2 + |h\nabla_x u|^2\right) dS_x \leq C\left(\frac{-\operatorname{Im} z_0(h)}{h} + e^{-\gamma(\rho)/h}\right) \int_{B(0,\rho)} |u|^2 dx, \qquad (42)$$

for any $\rho > R'_0$. Moreover, $\gamma(\rho) \to \infty$, as $\rho \to \infty$. More precisely, the analysis in [St3] shows that $\gamma \geq (\rho - R'_0)/C_0$. The constant $C$ above depends on $\rho$ but can be chosen locally uniform. Let $\chi \in C_0^\infty$, $\mathbf{1}_{B(0,R'_0)} \prec \chi \prec \mathbf{1}_{B(0,R_1)}$ Set
$$v(h) = \chi u(h).$$
Then $(P(h) - z_0(h))v(h) = [P_0(h), \chi]u(h)$. Note that by (42),

$$h\|\nabla\chi \cdot h\nabla u(h)\| + h^2 \|(\Delta\chi)u(h)\| = O\left(h^{1/2}\sqrt{-\operatorname{Im} z_0(h)} + e^{-\gamma(\rho)/h}\right) \|u(h)\|_{L^2(B(0,R_1))}. \qquad (43)$$

Therefore,

$$\|(P(h) - z_0(h))v(h)\| \leq C \left(h^{1/2} \sqrt{-\operatorname{Im} z_0(h)} + e^{-\gamma(\rho)/h}\right) \|u(h)\|_{L^2(B(0,R_1))}, \tag{44}$$

where $\gamma(\rho)$ has the properties above. Estimate (42) implies that $\|u(h)\|_{L^2(B(0,R_1))} \leq C\|v(h)\|$. We can replace $z_0(h)$ with $\operatorname{Re} z_0(h)$ and estimate (44) still holds. We regard now $v(h)$ as a quasimode for $Q(h)$, notice that $P(h)v(h) = Q(h)v(h)$, and an application of Theorem 4 and the remark preceding it yields that there exists an eigenvalue $w(h)$ of $Q(h)$ in (9) with different $\gamma(\rho)$ and $C$ having the same properties as above.

*Proof of (b):* Now, assume that $R_1 \leq R'_0$, see Figure 1. Then (44) is still true with $\chi$ such that $\chi \in C_0^\infty$, $\mathbf{1}_{B(0,R'_0)} \prec \chi \prec \mathbf{1}_{B(0,R'_1)}$ with $R'_1 > R'_0$, and $R_1$ in (44) is replaced by $R'_1$. Fix such a $\chi$. We will use semiclassical propagation of singularities argument to show that $v(h)$ is "small" not only outside $B(0, R'_0)$ as (42), (43) indicate but also outside $B(0, R_0)$. This is possible to do because $P(h)$ is non-trapping outside $B(0, R_0)$ for energy levels in $[a_0, b_0]$. We will use the propagation of singularities argument in the form presented in [St4, Lemma 4.1], see also [I]. A slight and obvious modification of the proof there implies that if $(P(h) - z_0(h))v(h) = g(h)$, where $\|g\|$ is bounded by the r.h.s. of (44)), and $v$ satisfies (42), then for any $\mu > 0$,

$$\|v(h)\|_{H^1(\mathbf{R}^n \setminus B(0,R_0+\mu))} \leq Ch^{-1/2} \left(\sqrt{-\operatorname{Im} z_0(h)} + O(h^\infty)\right) \|u(h)\|_{L^2(B(0,R'_1))}.$$

Now, we argue as in (a) to complete the proof of (b) by using the estimate above instead of (43).

*Proof of (c):* Let $w_0(h)$ be an eigenvalue of $Q(h)$ in (8) with eigenfunction $f(h)$, $\|f(h)\| = 1$. By (21),

$$\|(\operatorname{Re} W)^{1/2} f(h)\| = \sqrt{-\operatorname{Im} w_0(h)}. \tag{45}$$

Let $\chi \in C_0^\infty$ be such that $\mathbf{1}_{B(0,R_2)} \prec \chi \leq 1$, and consider $\chi f(h)$. Then

$$(P(h) - w_0(h))\chi f(h) = [P_0(h), \chi] f(h) + i\chi W f(h). \tag{46}$$

The latter term is $O(\sqrt{-\operatorname{Im} w_0(h)})$ by (45) and (5). Using standard semi-classical elliptic estimates, we get that $\|[-h^2\Delta, \chi] f(h)\| \leq Ch \|\mathbf{1}_{B(0,R_3)\setminus B(0,R_2)} f(h)\|$, where $R_3 \gg 0$. Using (45) again, we get that the latter is bounded by $C\sqrt{-\operatorname{Im} w_0(h)}$, because $\operatorname{Re} W \geq \delta_0$ for $|x| \geq R_2$ by (4). Thus,

$$\|(P(h) - w_0(h))\chi f(h)\| \leq C\sqrt{-\operatorname{Im} w_0(h)}. \tag{47}$$

By (45) and (4), $\|\chi f(h)\| \geq 1 - \sqrt{-\operatorname{Im} w_0(h)/\delta_0} \geq 1/2$, for $h$ small enough, so $\chi f(h)/\|\chi f(h)\|$ is a quasimode for $P(h)$. Applying Theorem 3, we get that there exists a resonance in (10). This completes the proof. $\square$

## 7 Proof of Theorem 2

The main arguments in this section are adapted from [St4], sections 3.3–3.5. As in the preceding section, we prove that cut-off resonant states of $P(h)$ are quasimodes of $Q(h)$; and cut-off eigenfunctions of $Q(h)$ are quasimodes of $P(h)$. To preserve the multiplicities and account for clusters of resonances too close to each other, we express $\Omega(h)$ as a union $\cup \Omega_j(h)$ of non-intersecting subdomains with small widths, and apply Theorem 3/Theorem 4 to each of them, showing that $m_j(h)$ resonances (eigenvalues) of $P(h)$ ($Q(h)$) in $\Omega_j(h)$ imply existence of at least $m_j(h)$ eigenvalues (resonances) of $Q(h)$ ($P(h)$) in a larger domain $\tilde{\Omega}_j(h) \supset \Omega_j(h)$ as in (41). The domains $\tilde{\Omega}_j(h)$ overlap, however, so we are in danger of counting some resonances several times. The critical moment in this approach is to prove that this does not happen, and in fact, there are at least $m(h) = \sum m_j(h)$ eigenvalues (resonances) in $\cup \tilde{\Omega}_j(h)$. This is achieved by showing that the set of all $m(h)$ cut-off resonant states (eigenfunctions) satisfy the property (40).

First, we recall the absorption estimate in [St4, Proposition 3.1], see also [B2, Proposition 6.1]. Let $P_\theta(h)$ be the complex scaled Hamiltonian with the complex scaling is performed outside $B(0, R'_0)$. More precisely, for some $B > R'_0$, we choose an increasing smooth function $0 \leq \theta(r) \leq \theta_0 = \text{const.} \ll 1$, such that $\text{supp}\, \theta \subset [B, \infty]$, $\theta(r) = \theta_0$ for $r > B + \delta/2$, with some $\delta > 0$. Then $P_\theta(h)$ is obtained from $P(h)$ by performing formally the change $x = r\omega \mapsto re^{i\theta(r)}\omega$ in polar coordinates. We refer to [SjZ1], [Sj1] for more details. We showed in [St4, Proposition 3.1] that for $h \ll h_1$, with some $h_1$,

$$\int (\theta + r\theta')|h\partial_r u|^2 + \theta \left(|hr^{-1}\nabla_\omega u|^2 + |u|^2\right) dx$$
$$\leq -\text{Im}\left(e^{i\theta}(P_\theta(h) - z)u, u\right) + \left(-\text{Im}\, z + e^{-h^{-1/3}}\right)\|u\|, \quad (48)$$

for any $z$ with $\text{Re}\, z \geq a_0$, $\text{Im}\, z \leq 0$, and $C = \min(a_0, 1)/2$. Observe that the requirement $B \gg 1$ in [St4] is not needed for compactly supported perturbations of the Laplacian that we study. In a remark following this proposition, it is claimed that one can replace $e^{-h^{-1/3}}$ there by $e^{-h^{-2/3+\varepsilon}}$, $\forall \varepsilon > 0$ with $h_1 = h_1(\varepsilon)$, if $\theta(r)$ is properly chosen.

To prove this, we will review the proof of (48), given in [St4, Proposition 3.1]. It is shown there that

$$-\text{Im}\left(e^{i\theta}(\tilde{P}_\theta - z)u, u\right) = I_1 + I_2 + I_3,$$

where, for $h \ll 1$,

$$I_1 \geq \frac{3}{4} \int \left((\theta + 2r\theta')|h\partial_r u|^2 + \theta|hr^{-1}\nabla_\omega u|^2\right) dx, \quad (49)$$

$$I_2 \geq \frac{3}{4} a_0 \int \theta |u|^2 dx, \quad (50)$$

and

$$I_3 = -\text{Im}\, h((\text{Re}\, g)h\partial_r u, u) + \frac{h^2}{2}(\text{Im}\, g'u, u) = I_3^{(1)} + I_3^{(2)}, \quad (51)$$

with

$$g(r) = \frac{d}{dr}\left(\frac{1}{1+ir\theta'}\right)\frac{e^{-i\theta}}{1+ir\theta'} = \frac{-i(r\theta'' + \theta')e^{-i\theta}}{(1+ir\theta')^3}. \quad (52)$$

Choose $\theta(r) = \exp(-(r-B)^{-k})$ for $0 \leq r \leq 1/C$, $C \gg 1$, $k > 0$. The function $g$ admits the following estimates

$$|\text{Re}\, g| \leq C(\theta' + |\theta''|)(\theta + \theta') \leq C\theta, \quad (53)$$
$$|g'| \leq C(\theta' + |\theta''| + |\theta'''|). \quad (54)$$

Now (53) implies that $I_3^{(1)}$ can be estimated by

$$|I_3^{(1)}| \leq Ch \int \theta(|h\partial_r u|^2 + |u|^2)\, dr\, d\omega \quad (55)$$

and for $h \ll 1$ this can be absorbed by the r.h.s. of (49) and (50). Next, to estimate $I_3^{(2)}$, we show that $\forall C \gg 0, \forall \varepsilon > 0, \exists k > 0$, such that

$$\theta' + |\theta''| + |\theta'''| \le h^{-2}\theta/C + e^{-h^{-2/3+\varepsilon}}$$

if $0 < h \ll 1$. The proof of the inequality above is done by considering two cases: $0 \le r - B \le h^{2/(k+3)}$, and $r - B \ge h^{2/(k+3)}$. Using this estimate, we see that $I_3^{(2)}$ can be absorbed by the r.h.s. of (50) as well. This completes the proof of (48) and explains the lower bound on $c(h)$ in Theorem 2.

*Proof of $N_P(\Omega(h)) \le N_Q(\Omega_+(h))$:* Fix $\varepsilon_0 > 0$ and let $\Omega(h)$ be as in (12) with $M \gg 1$ that will be determined later. We can assume that there are no resonances on $\partial\Omega(h)$. Let

$$\Omega_j(h) = [a_j(h), b_j(h)] + \mathrm{i}[-c(h), 0], \quad j = 1, \ldots, J(h) = O(h^{-n^\sharp}),$$

be non-intersecting domains such that all resonances in $\Omega(h)$ lie in the interior of some $\Omega_j(h)$. One can arrange the properties, as a consequence of (14)

$$\mathrm{dist}\big(\Omega_j(h), \Omega_k(h)\big) \ge 4w(h), \quad 0 < b_j(h) - a_j(h) \le Ch^{-n^\sharp}w(h), \tag{56}$$

where $0 < w(h) = O(h^N)$, $N \gg 1$ is fixed in advance. It is convenient to assume that (see [St4, Proposition 3.4])

$$w(h) = h^{-(5n^\sharp+1)/2}c(h). \tag{57}$$

Set

$$\Pi_{\Omega_j(h)} = \frac{1}{2\pi\mathrm{i}} \oint_{\partial\Omega_j(h)} (z - P_\theta(h))^{-1}\, \mathrm{d}z, \quad \Pi_{\Omega(h)} = \sum \Pi_{\Omega_j(h)}, \quad \mathcal{H}_\Omega = \Pi_{\Omega_j(h)}\mathcal{H}.$$

By [St4, Proposition 3.3], for all $j$,

$$\|(P_\theta(h) - a_j(h))u_j(h)\| \le Ch^{-6n^\sharp-1}c(h)\|u_j(h)\|, \quad \forall u_j(h) \in \Pi_{\Omega_j(h)}\mathcal{H}. \tag{58}$$

Furthermore, by [St4, Proposition 3.4], for all $j$,

$$\|\Pi_{\Omega_j(h)}\|_{\mathcal{H}_\Omega} \le Ch^{-(7n^\sharp+1)/2}. \tag{59}$$

This bound is the critical part of the proof that guarantees the property (40), as shown below.

Let $\chi_B$ be a smooth cut-off function such that $\mathbf{1}_{B(0,B+3\delta/4)} \prec \chi_B \prec \mathbf{1}_{B(0,B+\delta)}$. Then, by [St4, Theorem 3.1], for any collection of normalized $u_j(h)$ as above,

$$\|(P(h) - a_j(h))\chi_B u_j(h)\| + \|u_j(h) - \chi_B u_j(h)\| \le Ch^{-(3n^\sharp+1/2)}\sqrt{c(h)}. \tag{60}$$

Assume that $R'_0 < R_1$, i.e., $P(h) = -h^2\Delta$ near $\mathrm{supp}\, W$. Then we choose $B, \delta$ above to satisfy $R'_0 < B < R_1$, and $0 < 2\delta < R_1 - R'_0$. We now consider $\chi_B u_j(h)$ as quasimodes for $Q(h)$. At this point, we are mimicking the proof of [St4, Theorem 3.2] in the more difficult situation when the reference operator $Q(h)$ is not self-adjoint. For each $j = 1, \ldots, J(h)$, let $u_{jk}(h)$, $k = 1, \ldots, N_P(\Omega_j(h))$ be an orthonormal system in $\Pi_{\Omega_j}\mathcal{H}$. By the non-selfadjoint spectral theory, $u_{jk}(h)$ are linearly independent. It is the property (59) however, guaranteeing that this is preserved under small perturbations, that is needed in this proof. By (60), $\chi_B u_{jk}$ are quasimodes for $Q(h)$ as well, because $P_\theta(h)\chi_B = Q(h)\chi_B$. To verify (40), let $\tilde{u}_{jk}$ be another set of functions such that $\|\tilde{u}_{jk} - \chi_B u_{jk}\| \leq Ch^K$. Suppose that $\{\tilde{u}_{jk}\}$ are linearly dependent. Then
$$\sum c_{jk}\tilde{u}_{jk} = 0,$$
and we can assume that $\max_{jk} |c_{jk}| = 1$. Use (60) and the assumption on $\tilde{u}_{jk}$ above to get
$$\sum c_{jk} u_{jk} = O(h^{-n^\sharp})\left(h^K + h^{-(3n^\sharp+1/2)} h^{M/2}\right).$$
Let $j_0$ be the index for which $|c_{j_0 k_0}| = 1$ for some $k_0$. Apply $\Pi_{\Omega_{j_0}(h)}$ above, use (59), and the fact that the $u_{jk}$'s are orthonormal for a fixed $j$ to get
$$1 \leq \|\sum c_{j_0 k} u_{j_0 k}\| = O(h^{-(9n^\sharp+1)/2})\left(h^K + h^{-(3n^\sharp+1/2)} h^{M/2}\right).$$
We get a contradiction, if $K > (9n^\sharp + 1)/2$, $M/2 > 6n^\sharp + 1$. An application of Theorem 4 completes the proof of the estimate $N_P(\Omega(h)) \leq N_Q(\Omega_+(h))$ in case (a), i.e., when $R'_0 < R_1$.

Assume now that $R_1 \leq R'_0$. Then, as in the proof of Theorem 1(b), we propagate estimate (60) all the way to $\mathbf{R}^n \setminus B(0, R_0)$ at the expense of adding an $O(h^\infty)$ term. More precisely, for any $\chi \in C_0^\infty$ with $\mathbf{1}_{B(0,R_0)} \prec \chi$, we have the following
$$\|(P(h) - a_j(h))\chi u_j(h)\| + \|u_j(h) - \chi u_j(h)\| \leq Ch^{-(3n^\sharp+3/2)} \sqrt{c(h)} + O(h^\infty).$$
We now complete the proof in case (b) as above.

*Proof of $N_Q(\Omega_-(h)) \leq N_P(\Omega(h))$:* Similarly to the proof above, we show as in the preceeding section, that the cut-off eigenfunctions of $Q(h)$ are quasimodes of $P(h)$.

We will first estimate $N_Q(\Omega(h))$ from above. Let $\Omega_j(h)$ be as above, such that all eigenvalues of $Q(h)$ are included in the interior of some $\Omega_j(h)$, and there are no eigenvalues on $\partial \Omega(h)$. This decomposition can be done because of Proposition 2 and Proposition 4.

Since $Q(h)$ is non-selfadjoint, the multiplicity of each eigenvalue is the dimension of the span of the eigenvectors and the generalized eigenvectors (such that $(Q(h) - w(h))^k v = 0$ for some $k$). It is also given by the rank of the (non-orthogonal) spectral projection. Denote
$$\Pi_{\Omega_j(h)} = \frac{1}{2\pi i} \oint_{\partial \Omega_j(h)} (z - Q(h))^{-1} \, dz, \quad \Pi_{\Omega(h)} = \sum \Pi_{\Omega_j(h)}, \quad \mathcal{H}_\Omega = \Pi_{\Omega_j(h)} \mathcal{H}.$$
If $Q(h) = Q_R(h)$, then $\mathcal{H}$ above has to be replaced by $\mathcal{H}_R$. Proposition 3.3 in [St4] applies to $Q(h)$ as well, thanks to Propositions 2 and 3, thus (58) is true for $Q(h)$. Similarly, (59) holds as well.

Let $\|u_j(h)\| = 1$, $u_j(h) \in \Pi_{\Omega_j(h)} \mathcal{H}$. Then
$$\|(\operatorname{Re} W)^{1/2} u_j\|^2 = -\operatorname{Im}((Q - a_j)u_j, u_j) \leq Ch^{-(6n^\sharp+1)} c(h) \tag{61}$$
Let $\chi$ be as in section 6, and consider $\chi u_j$. Similarly to (46),
$$(P - a_j)\chi u_j = [P_0(h), \chi] u_j + i\chi W u_j + O\left(h^{-(6n^\sharp+1)} c(h)\right).$$
Following the arguments after (46), we get similarly to (60),
$$\|(P - a_j)\chi u_j\| + \|u_j - \chi u_j\| \leq Ch^{-(3n^\sharp+1/2)} \sqrt{c(h)}.$$
Then as above, we get $N_Q(\Omega(h)) \leq N_P(\Omega_+(h))$. To finish the proof, we set $\Omega_+(h) = \tilde{\Omega}(h) = [\tilde{a}, \tilde{b}] + i[-\tilde{c}, 0]$, and solve this for $a, b, c$. Note that in those arguments we only need to assume that $c(h) \geq e^{-C/h}$ since we are not using (48). $\square$